\documentclass[11pt,onecolumn]{article}
\usepackage{graphics,graphicx,epsfig,amsmath,color}
\usepackage[T1]{fontenc}
\usepackage[left=2cm,right=2cm,top=2cm,bottom=2cm,bindingoffset=0cm]{geometry}
\usepackage{subfig}
\usepackage{lmodern}
\usepackage{authblk}
\usepackage[english]{babel}

\begin{document}
\title{Spiral attractors as the root of a new type of ``bursting activity'' in the Rosenzweig-MacArthur model}
\author[1]{Yu. V. Bakhanova}
\author[1,2]{A. O. Kazakov}
\author[1]{A. G. Korotkov}
\author[1]{T. A. Levanova \thanks{tatiana.levanova@itmm.unn.ru}}
\author[1]{G. V. Osipov}
\affil[1,2]{Lobachevsky State University of Nizhny Novgorod, Institute of Information Technologies, Mathematics and Mechanics, 23, Prospekt Gagarina, Nizhny Novgorod, 603950, Russia}
\affil[2]{National Research University Higher School of Economics, 25/12 Bolshaya Pecherskaya Ulitsa, Nizhny Novgorod, 603155, Russia}
\maketitle
 
\abstract{We study the peculiarities of spiral attractors in the Rosenzweig-MacArthur model, that describes dynamics in a food chain ``prey-predator-superpredator''. It is well-known that spiral attractors having a ``teacup'' geometry are typical for this model at certain values of parameters for which the system can be considered as slow-fast system. We show that these attractors appear due to the Shilnikov scenario, the first step in which is associated with a supercritical Andronov-Hopf bifurcation and the last step leads to the appearance of a homoclinic attractor containing a homoclinic loop to a saddle-focus equilibrium with two-dimension unstable manifold. It is shown that the homoclinic spiral attractors together with the slow-fast behavior give rise to a new type of bursting activity in this system. Intervals of fast oscillations for such type of bursting alternate with slow motions of two types: small amplitude oscillations near a saddle-focus equilibrium and motions near a stable slow manifold of a fast subsystem. We demonstrate that such type of bursting activity can be either chaotic or regular.}

\section{Introduction}
Numerous biological experiments show that the dynamics in various life systems can be very diverse and complex. For example, even a model of a single neuron can demonstrate rather nontrivial dynamics related, in particular, to bursting activity -- periodic or chaotic oscillations for which sequences of fast oscillations (spikes) alternate with slow subthreshold motions \cite{Izhikevich2000}.

The basic property of the specified phenomenon is the existence in a system such type of orbit behavior, which is often called ``mixed mode oscillations'' \cite{GaspardKoper1992}, \cite{KuwamuraChiba2009}, \cite{GuckLis2015}. This type of motion is related to the existence of periodic or chaotic orbits with alternated oscillations of large and small amplitudes. One of the most interesting classes of systems demonstrating such type of orbit behavior are  systems with spiral attractors containing a homoclinic orbit (loop) to a saddle-focus equilibrium of type (1,2), i.e. with one-dimensional stable and two-dimensional unstable invariant manifolds. Recall, that the homoclinic loop is created when one of the branches (separatrices) of the stable manifold of a saddle-focus lies in the unstable  manifold. It follows from Shilnikov theorem \cite{Shilnikov1965} that the structure of the set of orbits located in a small neighborhood of the loop is chaotic. Moreover, it is naturally to select two different stages in such orbit behavior: the small amplitude stage, when the orbit passes near the equilibrium, and the large amplitude stage, when the orbit goes along a global piece of the homoclinic loop.

Note, that transition to the spiral chaos based on a saddle-focus (1,2) is realized typically in the accordance to the Shilnikov scenario \cite{Shilnikov1991}. This scenario can be observed in one parameter families (with control parameter $\mu$, for example) of three-dimensional flows, and its main steps, when varying (e.g. increasing) the value of $\mu$, are as follows. For $\mu < \mu_1$ an asymptotically stable equilibrium exists. At $\mu = \mu_1$ this equilibrium underdoes a supercritical Andronov-Hopf bifurcation: it becomes a saddle-focus (1,2), and a stable limit cycle $l$ is born. Then, at $\mu > \mu_2$, the limit cycle becomes of a focal type and a two-dimensional unstable manifold $W^u$ begins to wind up on it, forming a configuration resemble a whirlpool. This whirlpool tightens all trajectories from an absorbing domain except for one stable separatrix\footnote{The boundary of this whirlpool consists of the saddle-focus equilibrium and its two-dimension unstable invariant manifold having a form of cup, the edges of which are folded inside the whirlpool.}. With further increasing $\mu$ the cycle $l$ loses its stability, e.g. under a cascade of period doubling bifurcations\footnote{As it is observed in Rossler system \cite{Rossler1976}, Arneodo-Coullet-Tresser system \cite{ArnCoulTres1982}, Rosenzweig-MacArthur system \cite{RaiSreenivasan1993} etc.}, while the size of the whirlpool grows. Finally, at $\mu = \mu_3$, a homoclinic loop appears and a strange attractor, containing this loop occurs. We will call such attractors {\it homoclinic attractors}.

The general structure of the set of orbits on spiral attractors can be very different. The topology of such attractors is determined by the behavior of orbits passing through the boundary of the attractor (in other words, along the outer boundary of Shilnikov whirlpool). These orbits can make one, two, three or more turns before returning to the vicinity of a saddle-focus. In papers devoted to the study of attractors topology, e.g. \cite{Rossler1979}, \cite{ArgArnRich1987}, \cite{LetellierDutertreMaheu1995}, such attractors are often called screw, funnel and multi-funnel attractors, respectively (see Fig. \ref{fig:ScrewFunnel}). Thus, for screw attractors the large amplitude stage consists of one oscillation,  for funnel attractors this stage consists of two or three oscillations, and for multi-funnel attractors the large amplitude stage can have more than three oscillations.


Spiral attractors with different topology appear in such well-known models as Rossler system \cite{Rossler1976}, Rosenzweig-MacArthur system \cite{Rosenzweig1973}, \cite{HastingsPowell1991}, \cite{BakhanovaKazakovKorotkov2017},
Hindmarsch-Rose system \cite{HindmarshRose1984}, \cite{ShilnikovKolomiets2008}, chemical oscillator \cite{GaspardNicolis1983} etc.
Moreover, in all these cases the onset of chaos occurs due to the Shilnikov scenario.
It is worth noting that the above systems are, in fact, the slow-fast systems: the systems from \cite{HindmarshRose1984}, \cite{GaspardNicolis1983} are written in such a form where the slow and fast variables are separated explicitly, while in the systems from \cite{Rossler1976} and \cite{HastingsPowell1991} such separation can be performed at certain conditions on parameters, see e.g. \cite{MuratoriRinaldi1991}, \cite{KuznetsovRinaldi1996}. 

In the present paper we consider the Rosenzweig-MacArthur model when these conditions hold. We show that the mixed mode oscillations, occurring due to Shilnikov homoclinic orbit, together with the slow-fast behavior give rise to a new type of bursting activity in this system.
Intervals of fast oscillations for such type of bursting alternate with slow motions of two types: small amplitude oscillations near a saddle-focus equilibrium and motions near a stable slow manifold of a fast subsystem. We demonstrate that such type of bursting activity can be either chaotic or regular. Chaotic activity corresponds to multi-funnel strange attractors, while regular bursting corresponds to stable periodic orbits. Such periodic orbits are born near the homoclinic loop to a saddle-focus \cite{OvsyannikovShilnikov1986} and, therefore, they repeat the shape of this loop. Moreover, the appearance of periodic orbits, at the values of parameters from the so-called ``stability windows'', is an inalienable feature of quasiattractors \cite{AfrShil1983}, to which the spiral attractors of three-dimensional flows also belong.

\section{Equations}

This paper is devoted to the study of dynamics in the Rosenzweig-MacArthur model \cite{Rosenzweig1973} \cite{HastingsPowell1991}, describing the interaction of three different populations in tritrophic food chain which consists of a prey $x$, predator $y$, and superpredator $z$:
\begin{equation}
\begin{gathered}
\dot x = x\Big[r\Big(1-\frac{x}{K}\Big) - \frac{a_1 y}{1+b_1 x}\Big], \\
\dot y = y\Big[\frac{a_1 x}{1+b_1 x} - \frac{a_2 z}{1+b_2 y} - d_1\Big], \\
\dot z = z\Big[\frac{a_2 y}{1+b_2 y} -d_2 \Big].
\end{gathered}
\label{eq:mainEq}
\end{equation}
The detailed description of the model and its parameters can be found, for example, in \cite{HastingsPowell1991}. Here, following the paper \cite{KuznetsovDeFeoRinaldi2001}, we fix the parameters
\begin{equation}
a_1 = 5, a_2 = 0.1, b_1 = 3, b_2 = 2, d_1 = 0.4, d_2 = 0.01
\label{eq:params}
\end{equation}
and use $K$ and $r$ as control parameters.

\section{Shilnikov scenario and homoclinic attractors}

From many studies, e.g. \cite{RaiSreenivasan1993}, \cite{HastingsPowell1991}, \cite{KuznetsovRinaldi1996}, \cite{McCannYodzis1994}, it became clear that the dynamics in the Rosenzweig-MacArthur system can be chaotic. In \cite{RaiSreenivasan1993}, \cite{McCannYodzis1994} it was shown that strange attractors in this system can appear due to a cascade of period doubling bifurcations. On the other hand, as it was shown in the papers \cite{KuznetsovDeFeoRinaldi2001}, \cite{DengHines2002}, homoclinic attractors, that occur due to a homoclinic orbits to a saddle-focus equilibrium, can also appear in the system. In \cite{DengHines2002} the existence of such homoclinic orbits was established using multi-time scale analysis, while in \cite{KuznetsovDeFeoRinaldi2001} it was done using a numerical continuation technique.
\begin{figure}[!ht]
\centering
\subfloat[$K = 0.94$]
{
	\includegraphics[width = .3\linewidth]{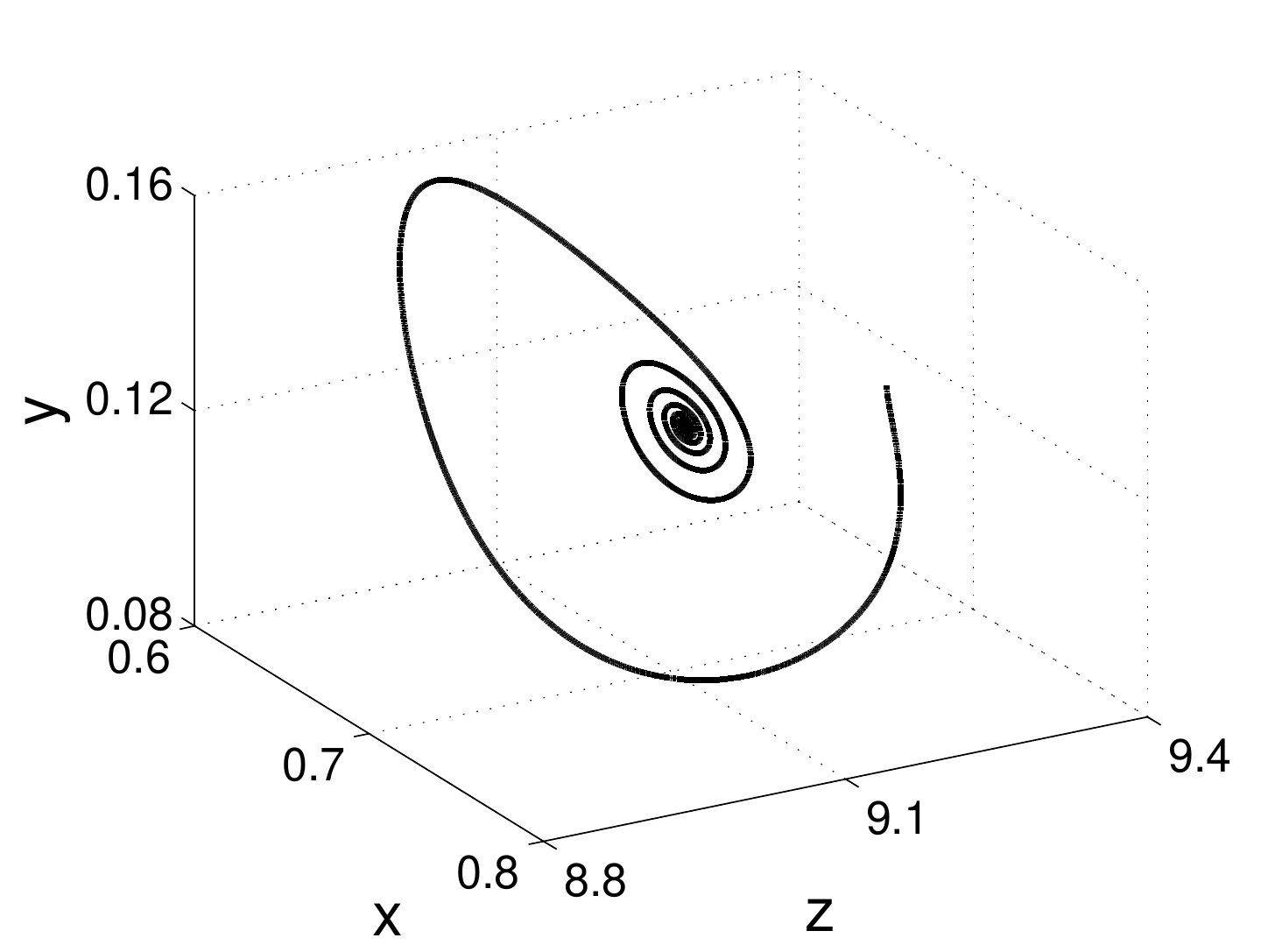}
	\label{fig:1a}
}
\subfloat[$K = 1$]
{
	\includegraphics[width = .3\linewidth]{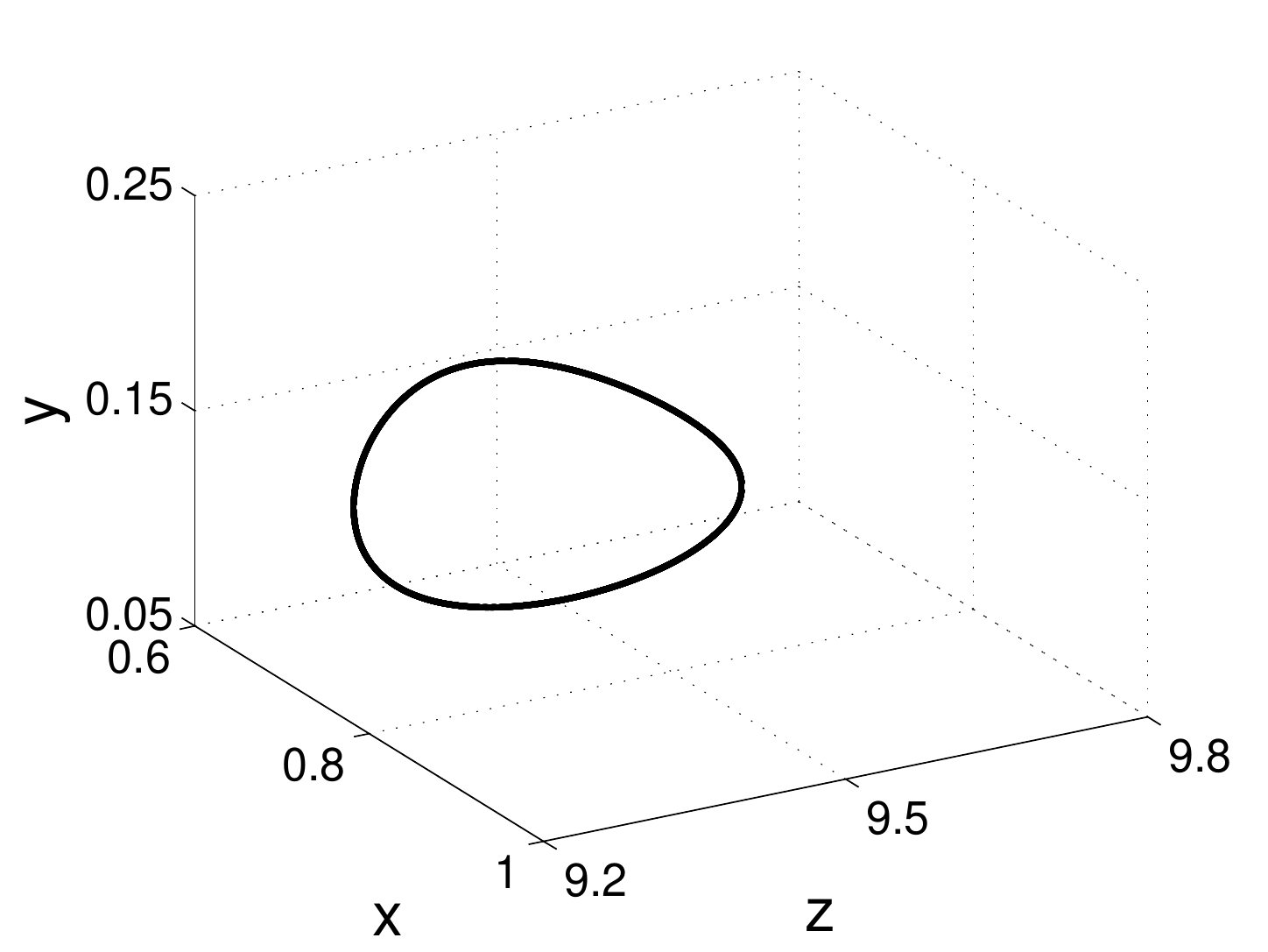}
	\label{fig:1b}
}
\subfloat[$K = 1.03$]
{
	\includegraphics[width = .3\linewidth]{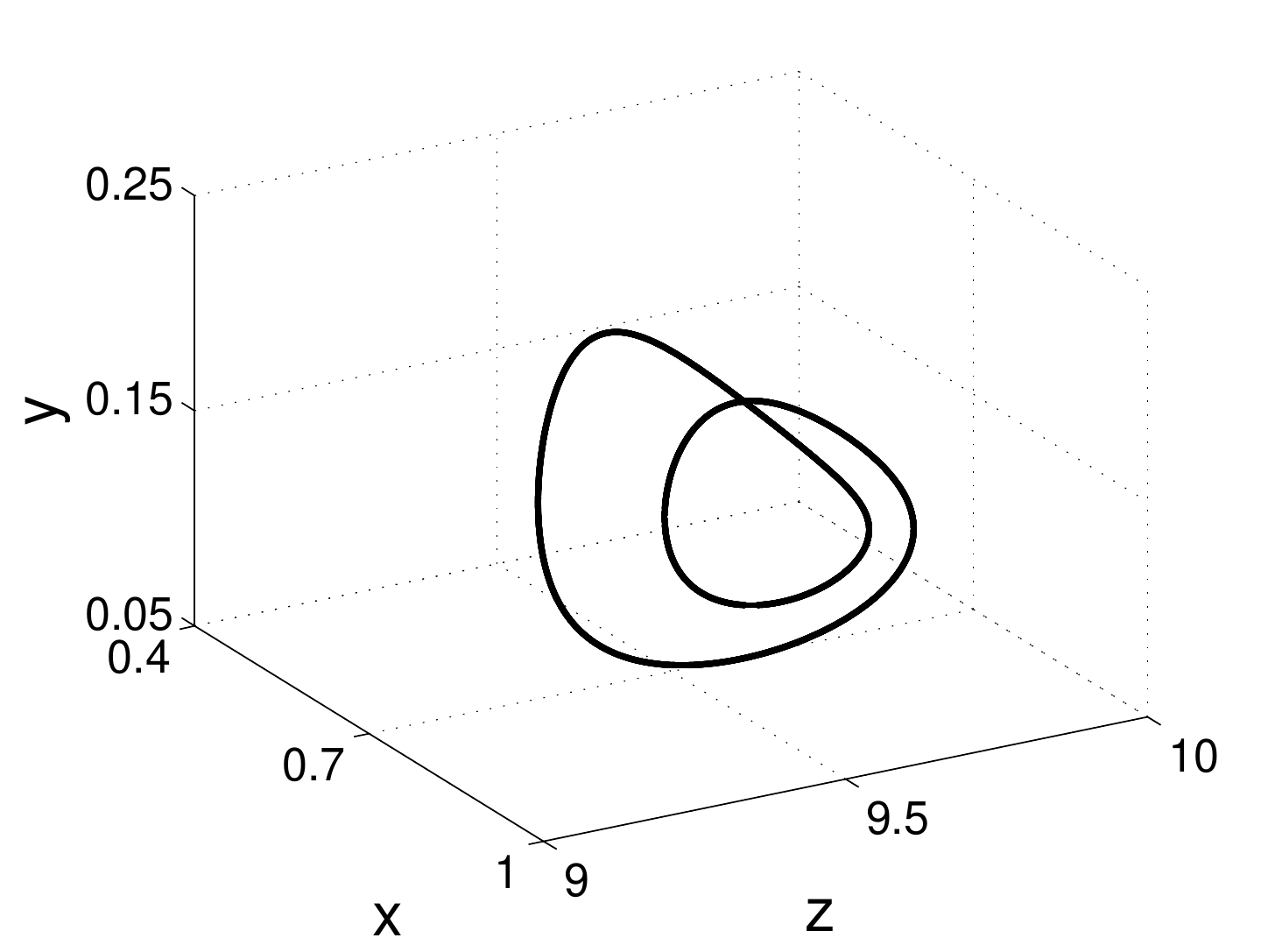}
	\label{fig:1c}
}\\
\subfloat[$K = 1.036$]
{
	\includegraphics[width = .3\linewidth]{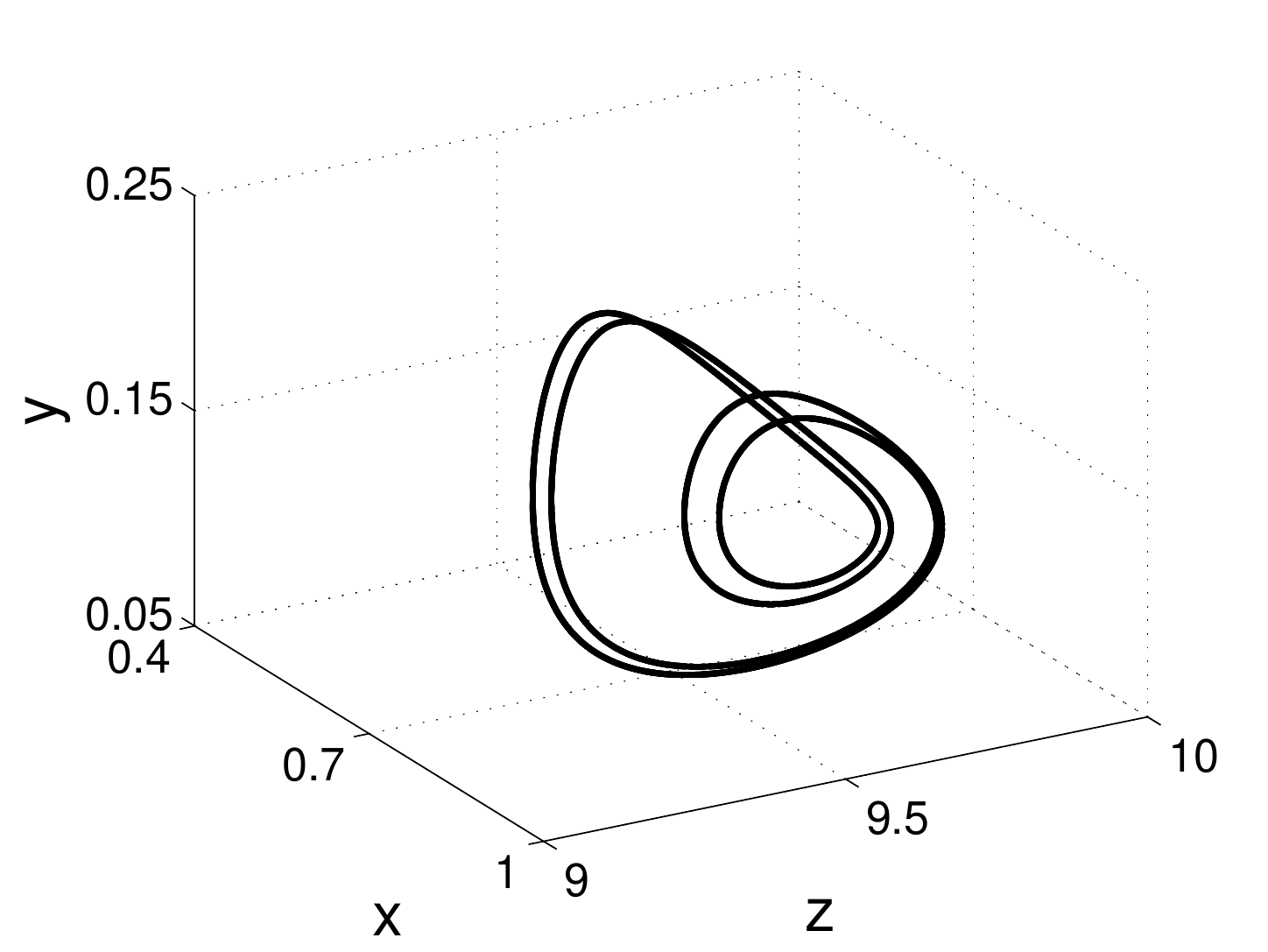}
	\label{fig:1d}
}
\subfloat[$K = 1.040115$]
{
	\includegraphics[width = .3\linewidth]{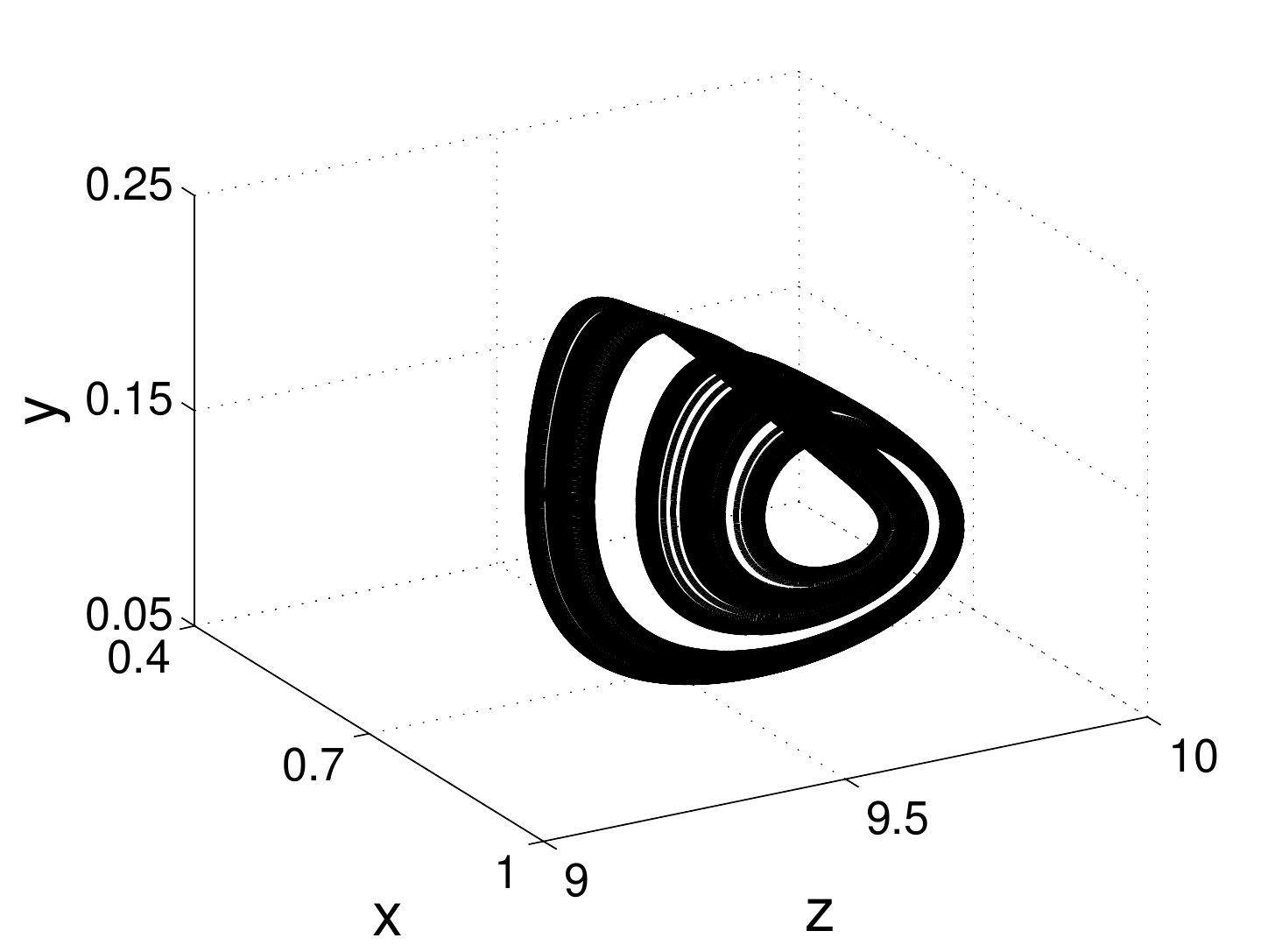}
	\label{fig:1e}
}
\subfloat[$K = 1.06356$]
{
	\includegraphics[width = .3\linewidth]{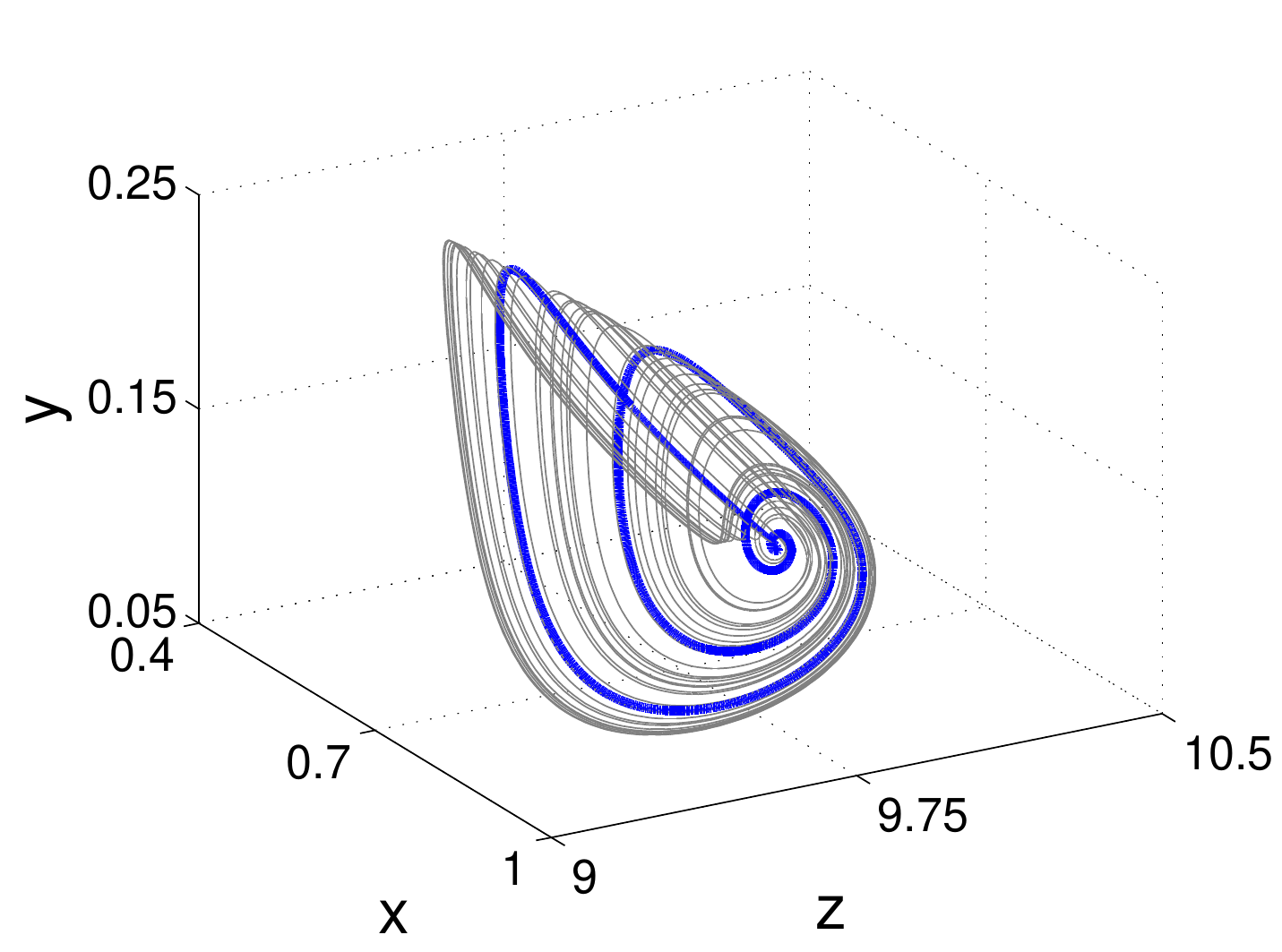}
	\label{fig:1f}
}
\caption{The illustration of the Shilnikov scenario: (a) stable equilibrium; (b) stable limit cycle; (c) 2-period stable limit cycle; (d) 4-period stable limit cycle; (e) attractor after a cascade of period doubling bifurcations; (f) homoclinic spiral attractor, blue bold curve is a homoclinic orbit, dark grey -- trajectory on the attractors.}
\label{fig:1}
\end{figure}

Further, we fix $r = 0.79$ and perform one-parameter analysis increasing the parameter $K$. When $K < K_1 \approx 0.965$, positive stable equilibrium is a global attractor in the system, see Fig. \ref{fig:1a}.  At $K = K_1$  this equilibrium undergoes Andronov-Hopf bifurcation due to which a stable limit cycle appears and the equilibrium becomes of a saddle-focus (1,2) type,  see Fig. \ref{fig:1b}. With further increase in the parameter $K$, limit cycle undergoes the cascade of period-doubling bifurcation (see cycles of 2- and 4- periods in Fig. \ref{fig:1c} and \ref{fig:1d}) and strange attractor of a Feigenbaum type occurs, see Fig. \ref{fig:1e}. At $K = K_h \approx 1.06356$ the homoclinic loop to the saddle-focus equilibrium appears and a homoclinic attractor is born, see Fig. \ref{fig:1f}.


Note, that the same scenario occurs with increasing of parameter $r$, when $K$ is fixed (for example, $K = 1.1$). Thus, a cascade of period doubling bifurcations is only the first part in the scenario of the development of chaotic dynamics in the system \eqref{eq:mainEq}. The final part of this scenario is the appearance of homoclinic orbits to the saddle-focus equilibrium of (1,2) type due to which attractors become of the homoclinic type, i.e. it can be argued, that the Shilnikov scenario is a typical scenario of the onset of chaos in the system under consideration.

\begin{figure}[!ht]
\centering
\begin{minipage}[h]{0.32\linewidth}
\center{\includegraphics[width=1\linewidth]{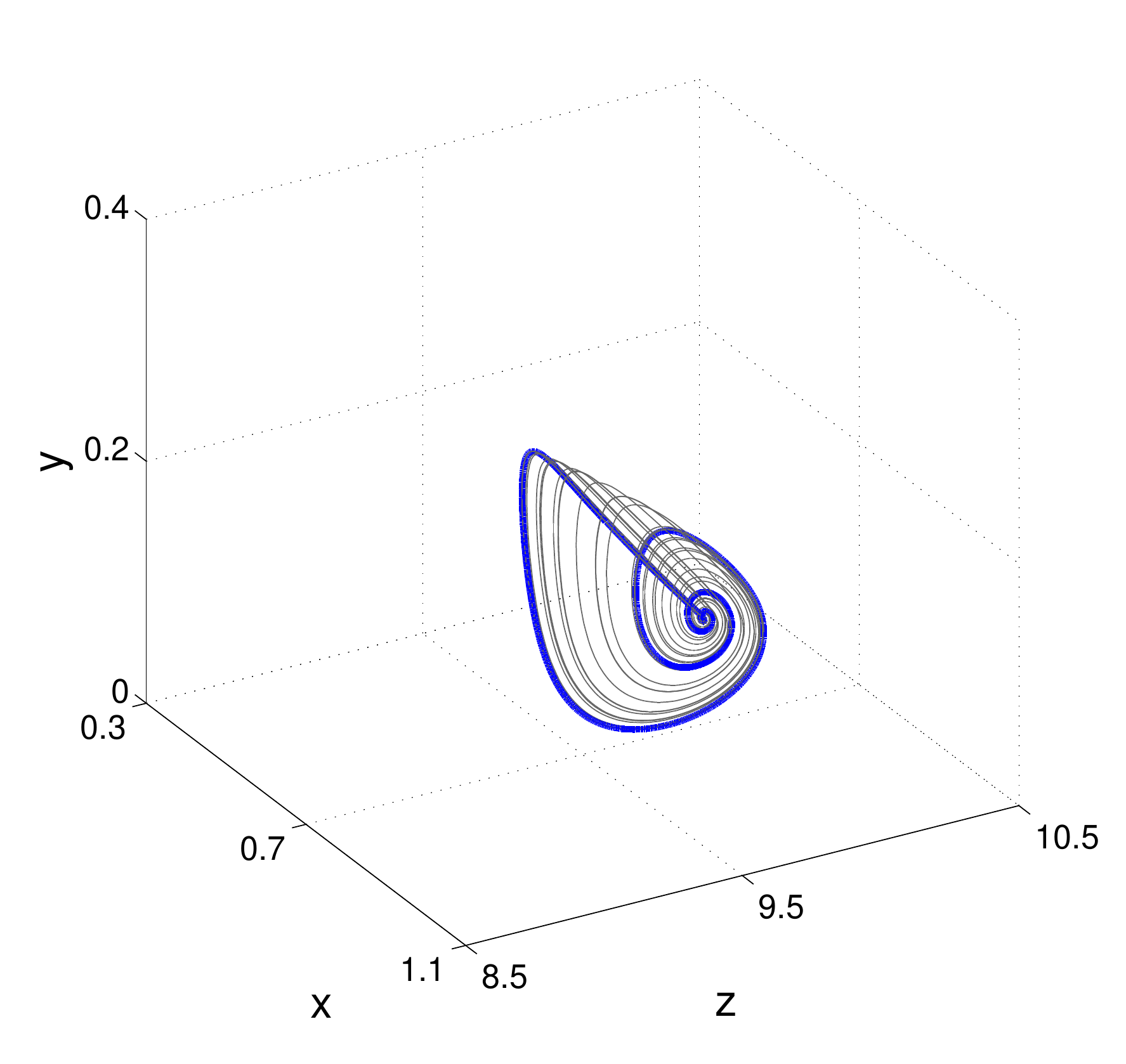} \\ a) {\footnotesize $r = 0.782, K = 1.059$}}
\end{minipage}
\hfill
\begin{minipage}[h]{0.32\linewidth}
\center{\includegraphics[width=1\linewidth]{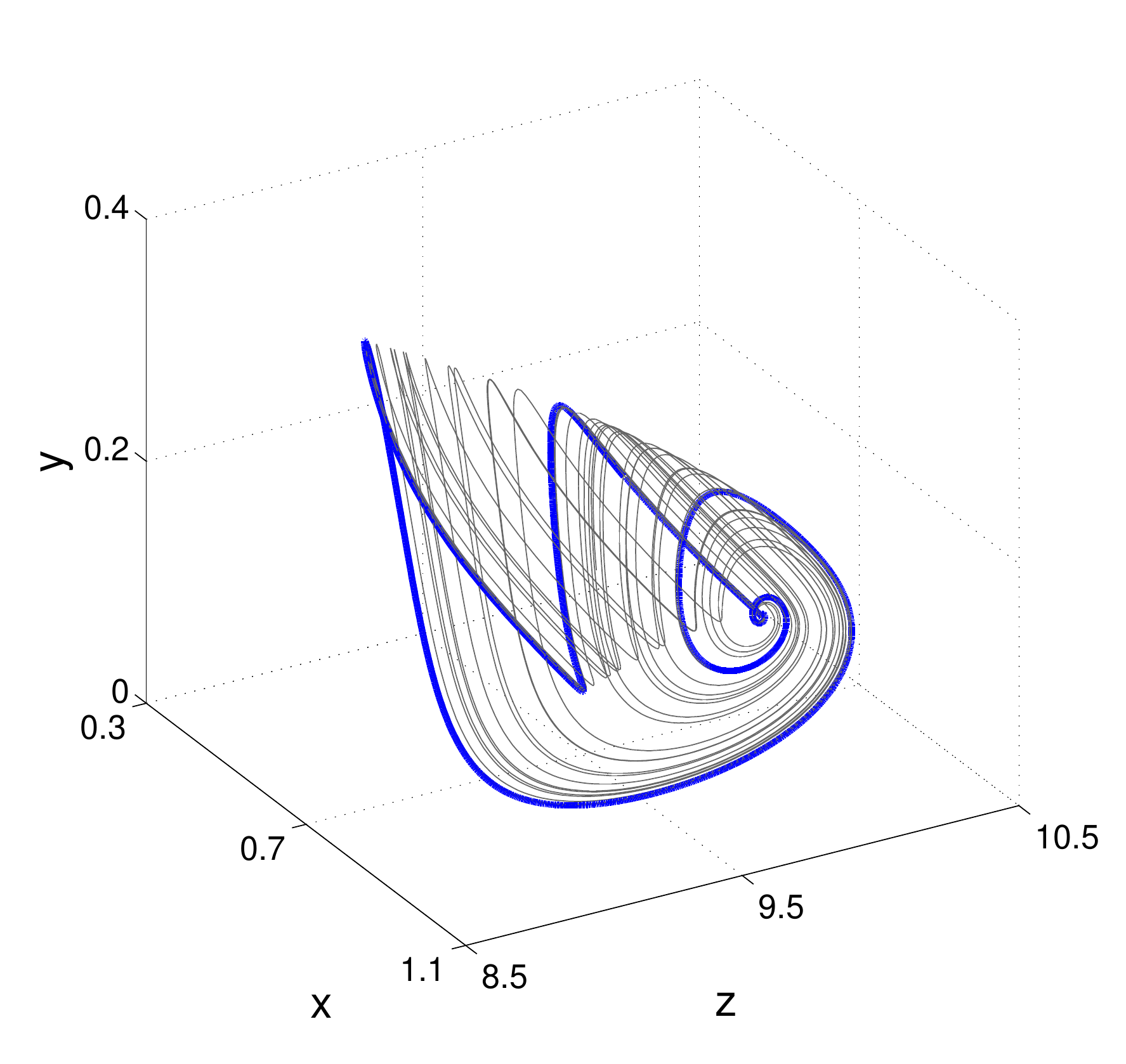} \\ b) {\footnotesize $r = 0.8122, K = 1.08$}}
\end{minipage}
\hfill
\begin{minipage}[h]{0.32\linewidth}
\center{\includegraphics[width=1\linewidth]{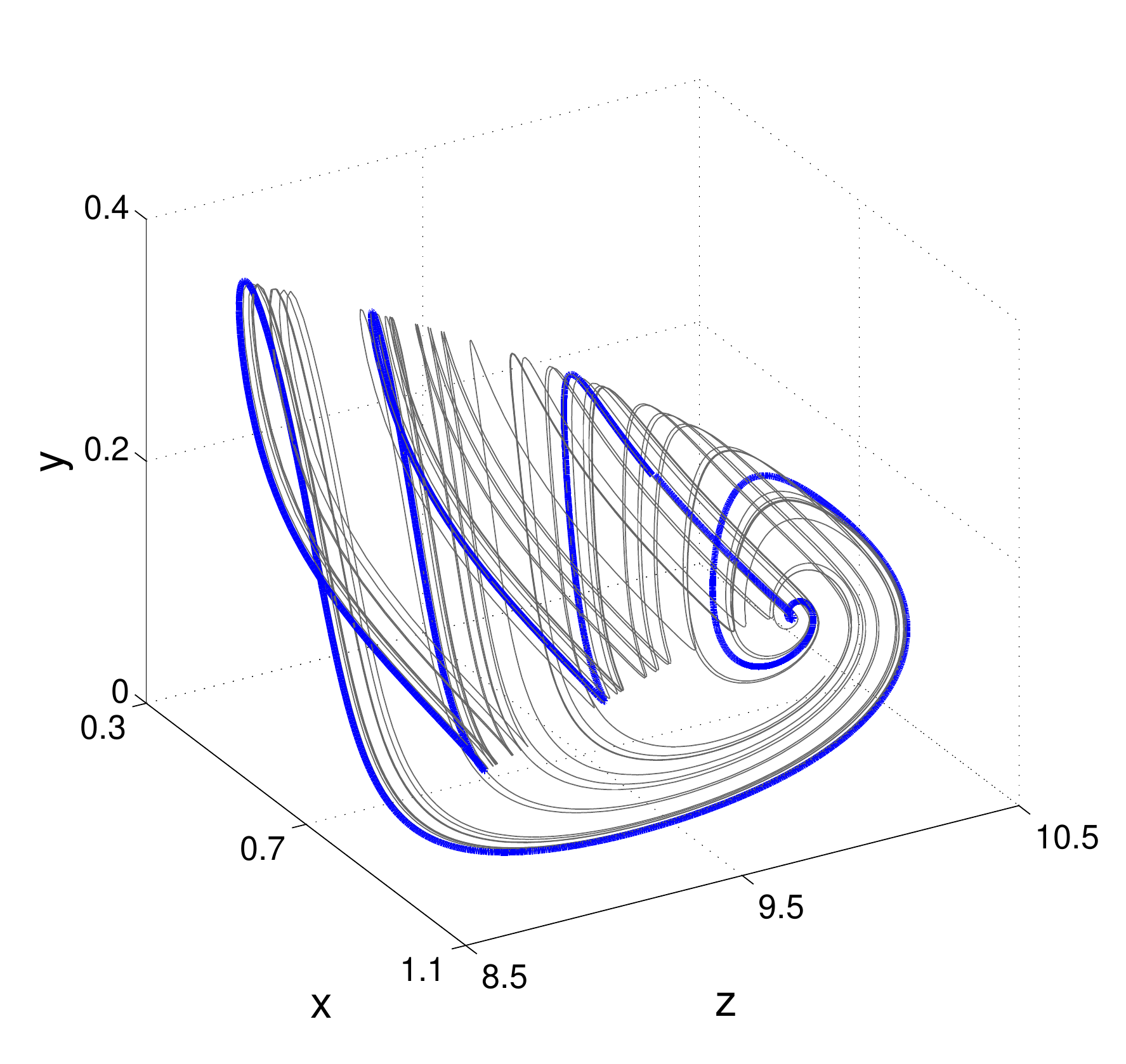} \\ c) {\footnotesize $r = 0.8356, K = 1.09$}}
\end{minipage}
\caption{{\footnotesize Different types of homoclinic attractors. (a) Screw homoclinic attractor; (b) funnel homoclinic attractor; (c) multi-funnel homoclinic attractor. The trajectory on attractors is colored in dark gray, blue bold curves are the homoclinic orbits.}}
\label{fig:ScrewFunnel}
\end{figure}

The examples of various homoclinic attractors in the system are presented in Fig. \ref{fig:ScrewFunnel}. These attractors differ by the complexity of orbits passing through the boundary of the attractors. In Fig. \ref{fig:ScrewFunnel}a such orbits make one turn before they come back into vicinity of the saddle-focus, while in Figs. \ref{fig:ScrewFunnel}b, \ref{fig:ScrewFunnel}c such orbits has two and three turns, respectively. In papers devoted to the study of topology of attractors, e.g. \cite{Rossler1979}, \cite{ArgArnRich1987}, \cite{LetellierDutertreMaheu1995}, the attractors from Fig. \ref{fig:ScrewFunnel} are often called screw, funnel and multi-funnel, respectively. Note, that multi-funnel homoclinic attractor in the system can have upon 9 turns in its funnel.

\section{Singular perturbation analysis}

Spiral attractors with different topology can be non-homoclinic. Apparently, the paper \cite{HastingsPowell1991} was the first work where in the Rosenzweig-MacArthur system non-homoclinic attractors with complex topology were discovered. In that paper such attractors were called  ``teacup'' attractors, see Fig. \ref{fig:chaotic_bursting}a. These attractors correspond to the irregular slow oscillations between predator and superpredator accompanied by several fast predator-prey oscillations on a ``period'' of slow oscillations, see Fig. \ref{fig:chaotic_bursting}b.

With varying of the parameters ``teacup'' attractors can change its topology: the number of turns in their funnel can increase or decrease. Such orbits behavior can be explained using singular perturbation analysis (SPA) which is based on a slow-fast decomposition of the system.

We note that in the case under consideration the parameters $a_2$ and $d_2$ in the equation for superpredator are an order of magnitude smaller than corresponding parameters $a_1$, $d_1$ and $r$ in the equations for both predator and prey. Therefore, the subsystem
\begin{equation}
\begin{gathered}
\dot x = x\Big[r\Big(1-\frac{x}{K}\Big) - \frac{a_1 y}{1+b_1 x}\Big], \\
\dot y = y\Big[\frac{a_1 x}{1+b_1 x} - \frac{a_2 z}{1+b_2 y} - d_1\Big]
\end{gathered}
\label{eq:fast_subsystem}
\end{equation}
consisting of equations for prey $x$ and predator $y$ can be considered as a fast subsystem of \eqref{eq:mainEq}. The details on the singular perturbation technics for different types of slow-fast systems can be found in the paper \cite{MuratoriRinaldi1991} where, in particular, such analysis was applied for the system with one slow and two fast variables (like in our case).
For the system under consideration SPA was described in the paper \cite{KuznetsovRinaldi1996}, where also the detailed bifurcation analysis of the system was performed on the plane $(d_1, d_2)$ and diagrams illustrating singular perturbation approach was constructed for different values of $d_1$.

Usually singular perturbation analysis for slow-fast systems is carried out in 2 main stages. In our case, at the first stage, the coordinate $z$ is considered as a parameter and all stable invariant manifolds (equilibria and limit cycles) in the fast subsystem are determined for different values of $z$. At the second stage, the equilibrium manifold of the slow subsystem given by $\dot z = 0$ condition is used for constructing singular orbits which approximate real trajectories of the system. For visual presentation of the results of SPA it is quite convenient to map all invariant sets (equilibria and limit cycles) to $(x, y, z)$ space. In such a representation equilibria are mapped to a curves, while limit cycles are maped to paraboloid-like surface.

\begin{figure}[!ht]
\center{\includegraphics[width=1.0\linewidth]{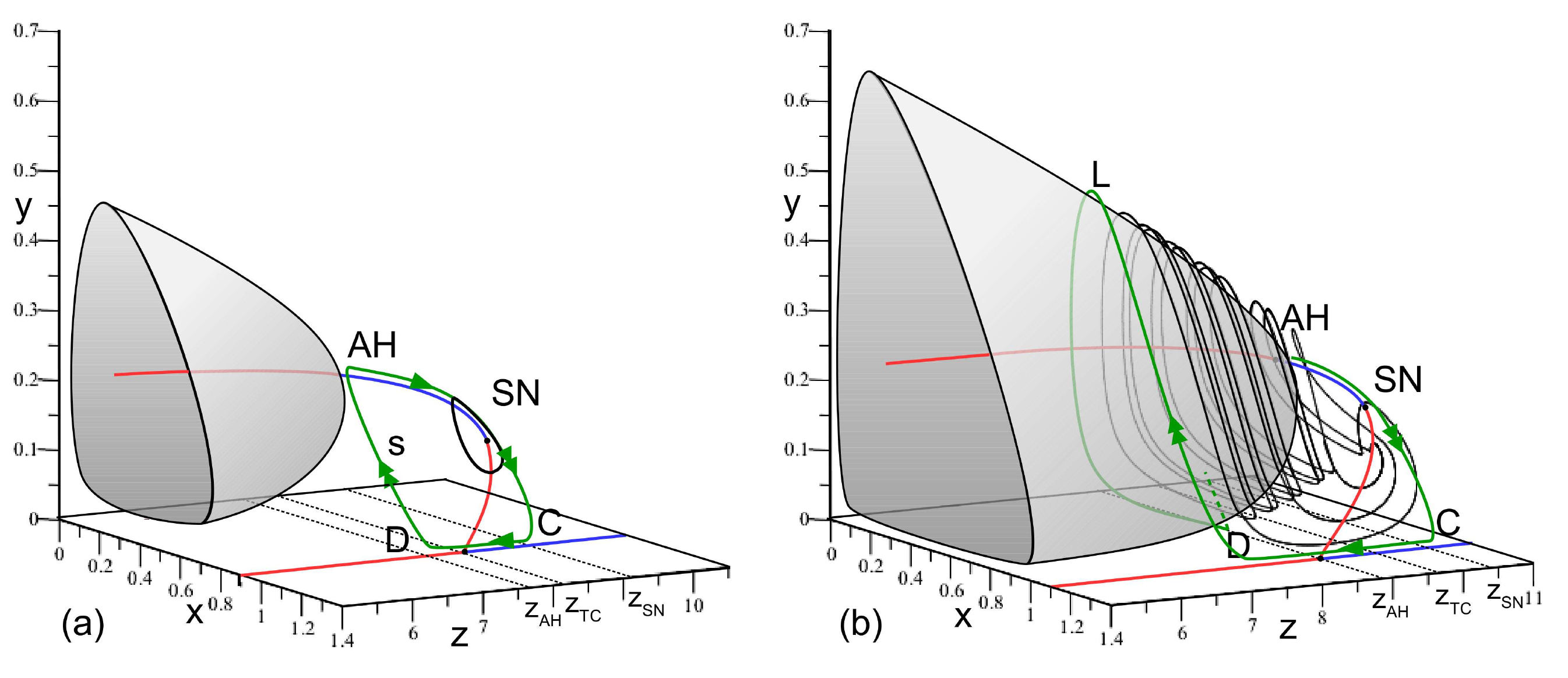} }
\caption{{\footnotesize Illustrations to singular perturbation analysis for $r = 0.9$. In blue color -- stable equilibria, in red color -- unstable or saddle equilibria in the fast subsystem \eqref{eq:fast_subsystem} for different values of $z$. In green color -- singular orbits, in black color -- real orbits of the system \eqref{eq:mainEq}. Green and black arrows indicate the direction on singular and real orbits. Single arrows correspond to slow motions along stable manifolds, while double arrows correspond to fast jumps or oscillations. (a) K = 0.9; (b) K = 1.1.}}
\label{fig:SPA}
\end{figure}

Here we fix the parameter $r = 0.9$ and perform SPA for $K = 0.9$ and $K = 1.1$. Fig. \ref{fig:SPA} shows equilibria and stable limit cycles of the fast subsystem \eqref{eq:fast_subsystem} depending on the parameter $z$. In the both cases ($K = 0.9$ and $K = 1.1$) for large values of $z$ there is only one stable equilibrium $e_0: (x, y) = (K, 0)$ which is located in the invariant plane $y = 0$. At $z = z_{SN}$ a saddle-node bifurcation occurs and a pair of stable $e_1$ and saddle $e_2$ equilibria appears at the positive quadrant (see blue and red branches outgoing from the point $SN$ in Fig. \ref{fig:SPA}). With further decrease of the parameter $z$ the saddle equilibrium tends to the invariant plane $y=0$ and, at $z = z_{TC}$ undergoes transcritical bifurcations due to which stable equilibrium $e_0$ becomes a saddle one for $z < z_{TC}$. Another principal bifurcation occurs with the stable equilibrium $e_1$ at $z = z_{AH} $, when this equilibrium loses its stability due to supercritical Andronov-Hopf bifurcation and, finally, a stable limit cycle appears for $z < z_{AH}$.


As it is known from SPA theory, the real trajectories of slow-fast systems are well approximated by singular orbits, which are composed from slow motions along the stable manifolds (equilibria or limit cycles) of the fast subsystem and fast jumps near subcritical bifurcations of stability loss. The direction of singular orbits along slow manifolds is determined by the sign of $\dot z$. Namely, from the third equation of \eqref{eq:mainEq} it follows, that if $y > y_0$, where
$$
y_{0} = \frac{d_2}{a_2-b_2 d_2} = 0.125,
$$
then $\dot z > 0$, otherwise $\dot z \leq 0$.

Let us describe the behavior of singular orbits for $K = 0.9$ and for $K = 1.1$. In the both cases the stable equilibrium $e_1$ is located above the plane $y_0 = 0.125$. Therefore, singular orbits tend to the right along the branch corresponding to $e_1$ (see $AH$-$SN$ segment in Fig. \ref{fig:SPA}). At the point $SN$ the stable equilibrium $e_1$ collides with the saddle $e_2$ and singular orbits jump to the stable equilibrium $e_0$ located on the plane $y = 0$ (segment $SN$-$C$ in Fig. \ref{fig:SPA}) intersecting the plane $y_0 = 0.125$. At the lower branch $C$-$D$, which corresponds to the equilibrium $e_0$, singular orbits go to the left (since $\dot z$ here is negative) until the point $D$, where these orbits jump back either to the branch corresponding to the stable equilibrium $e_1$ (like in Fig. \ref{fig:SPA}a) or to the paraboloid-like surface $L$ which corresponds to stable limit cycles (like in Fig. \ref{fig:SPA}b)\footnote{We note that singular orbits in Fig. \ref{fig:SPA} jump from the point $D$ to a point $A$ not immediately after $e_0$ equilibrium loses its stability. Such orbits behavior is explained by the Pontryagin's delay of stability \cite{Pontryagin1957}, \cite{MuratoriRinaldi1992}.}.

We emphasize that in the point $D$ two principally different cases are possible. In the first case, when, for example, $K = 0.9$, singular orbits jump back to the branch $AH$-$SN$ corresponding to the stable equilibrium $e_1$. In this case the appearance of multi-funnel attractor is impossible, moreover, a stable equilibrium or a limit cycle (maybe of a large period) are typical attractors for the full system here
(see black colored limit cycle in Fig. \ref{fig:SPA}a). In the second case, singular orbits jump to the paraboloid-like surface $L$ and move fast around this surface in a screw manner until Andronov-Hopf bifurcation occurs with $e_1$ equilibrium at the point $AH$, see Fig. \ref{fig:SPA}b.

\begin{figure}[t]
\center{\includegraphics[width=0.8\linewidth]{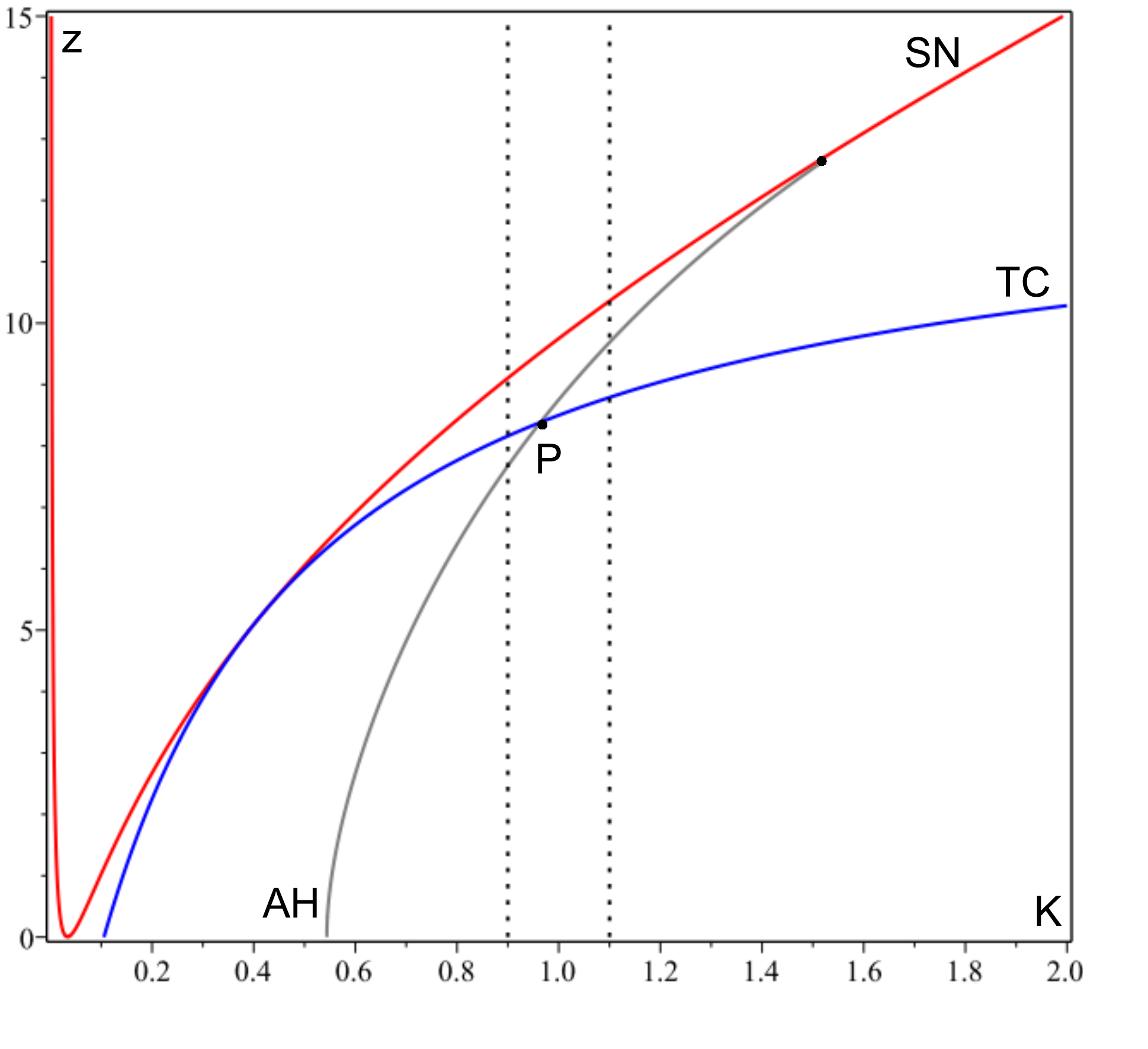} }
\caption{{\footnotesize Bifurcation diagram of the fast subsystem \eqref{eq:fast_subsystem}. {\bf SN} -- the curve of a saddle-node bifurcation. A pair of stable $e_1$ and saddle $e_2$ equilibria appears under this curve. {\bf AH} -- the curve of Andronov-Hopf bifurcation. The stable equilibrium $e_1$ loses its stability and a stable limit cycle $l$ appears under this curve. {\bf TC} -- the curve of a transcritical bifurcation. The positive saddle equilibrium $e_2$ becomes stable and negative under this curve, while stable equilibrium $e_0$ located on the plane $y=0$ becomes a saddle one.}}
\label{fig:FasSub_BD}
\end{figure}

We note that these two cases differ by the order of the occurrence of Andronov-Hopf and transcritical bifurcations for the fast subsystem \eqref{eq:fast_subsystem}. For convenience two parametric bifurcation diagram (on $(K, z)$ - plane) of the fast subsystem is presented in Fig. \ref{fig:FasSub_BD}.
Here {\bf SN} -- the curve of a saddle-node bifurcation, under which a pair of stable $e_1$ and saddle $e_2$ equilibria appears. {\bf AH} -- the curve of Andronov-Hopf bifurcation, where the stable equilibrium $e_1$ loses its stability and a stable limit cycle $l$ occurs. {\bf TC} -- the curve of transcritical bifurcation. The positive saddle equilibrium $e_2$ becomes stable and negative under this curve, while stable equilibrium $e_0$ located on the plane $y=0$ becomes a saddle one. From Fig. \ref{fig:FasSub_BD} it is clear that multi-funnel attractors (with many turn in the funnel) can appear here only for $K > K_p \approx 0.964$.

\begin{figure}[!ht]
\center{\includegraphics[width=1.0\linewidth]{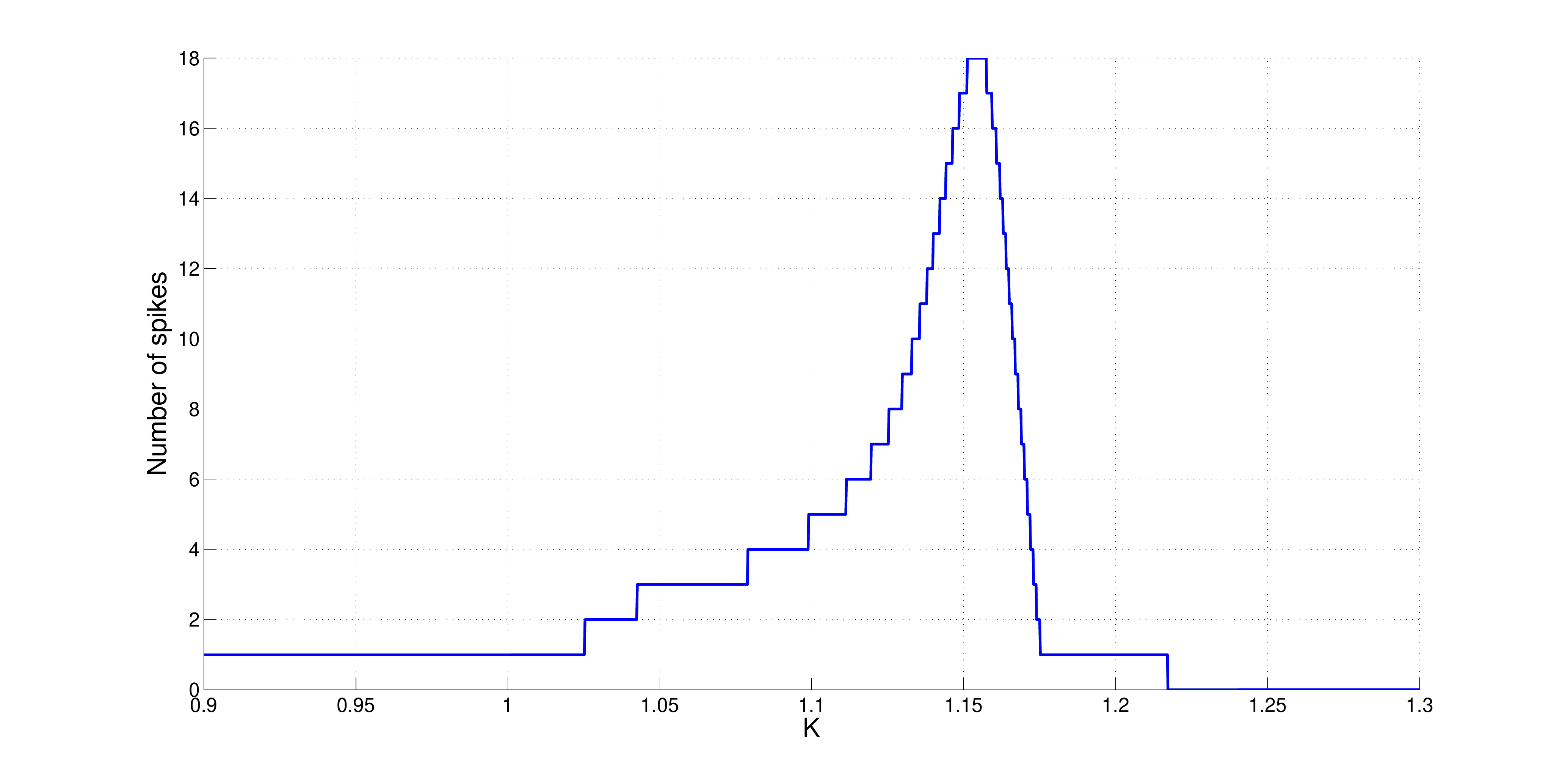} }
\caption{{\footnotesize The dependency of the number of turns in the funnel of the attractor from the parameter $K$. Here $r = 0.9$ and the other parameters are given by \eqref{eq:params}. Thus, non-homoclinic spiral attractors can have up to 18 turns in its funnel.}}
\label{fig:NumberOfTurns}
\end{figure}

It is interesting to understand how the number of turns in the funnel depends on the parameter $K$. For obtaining this dependency we use quite simple techniques. For each values of $K$ we calculate quite long time series $y(t)$ and $z(t)$ on the attractor and estimate the maximum number of peaks in bursts from $y(t)$, separating different bursts by $z$ coordinates from the series $z(t)$. If $z$ coordinate becomes greater than $z$ coordinate of saddle-focus equilibrium of the full system \eqref{eq:mainEq} then we suppose that the burst is over. The described dependency presented in Fig. \ref{fig:NumberOfTurns}, where one can see that first, with the increasing of $K$, the number of turns increases too. At $K \approx 1.16$ the number of turns reaches its maximum 18, after which this number decreases to 0 at $K > 1.195$, where the attractor is a stable limit cycle.

\section{New type of bursting activity in the model}

\begin{figure}[!ht]
\center{\includegraphics[width=0.9\linewidth]{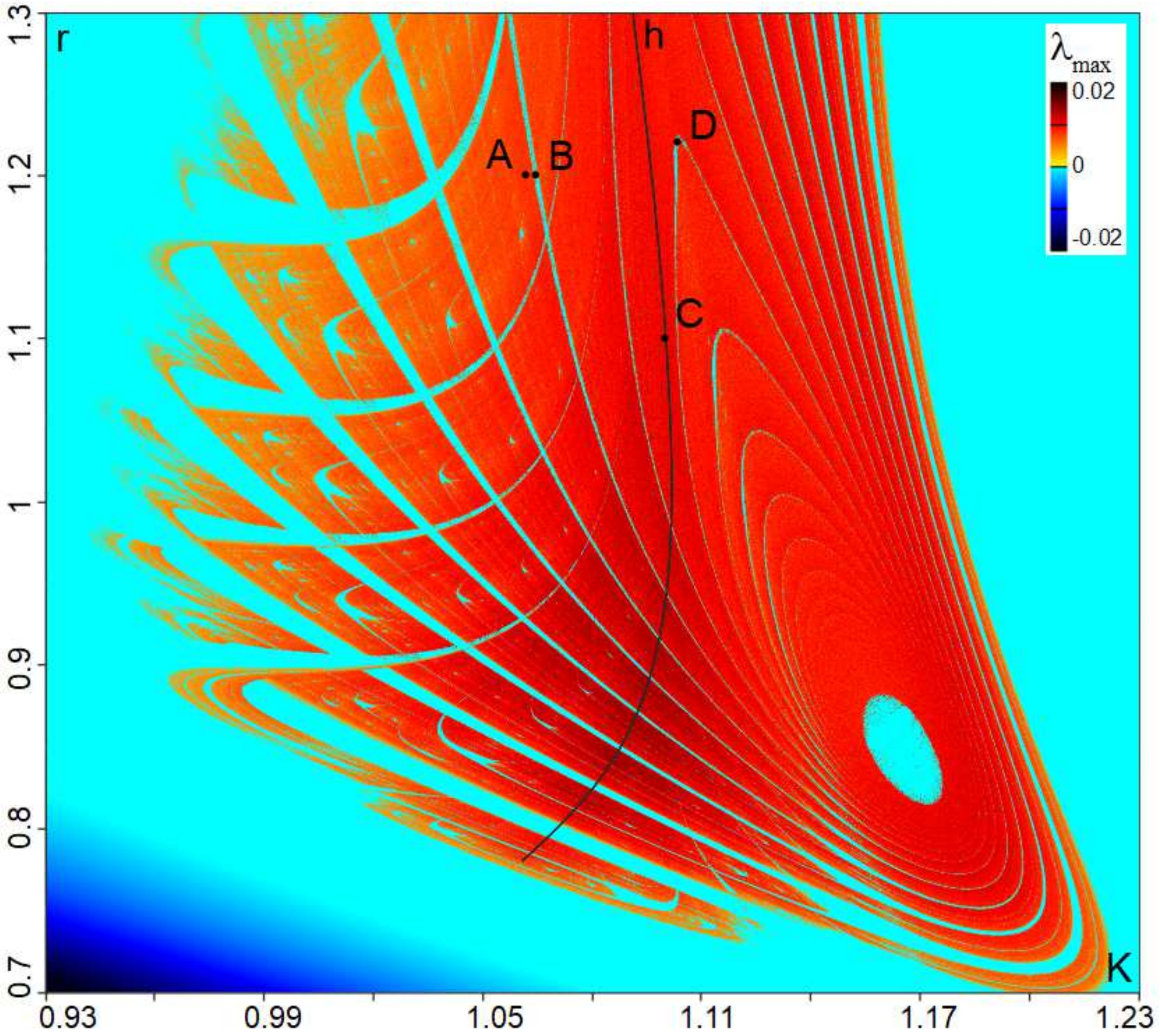} }
\caption{{\footnotesize Diagram of maximal Lyapunov exponent. Blue color correspond to stable periodic orbits, yellow and red colors correspond to chaotic regimes, and black colored line is the line of homoclinic bifurcation curve. The points $A$, $B$, $C$, and $D$ correspond to four different types of bursting activity.}}
\label{fig:LE_Diagram}
\end{figure}

Due to the slow-fast behavior and the existence of multi-funnel attractors the Rosenzweig-MacArthur system exhibits various types of bursting activity. Here we present the description of these types of activity and point out them on the diagram of maximal Lyapunov exponent, see Fig. \ref{fig:LE_Diagram}. In this figure shades of blue color correspond to stable periodic orbits, while shades of yellow and red colors correspond to chaotic regimes (see legend in the top right corner of Fig. \ref{fig:LE_Diagram}). Also we put on this diagram the homoclinic bifurcation curve $h$ (black colored line) and four points $A$, $B$, $C$, and $D$, which correspond to four different types of bursting activity.

\begin{figure}[!ht]
\center{\includegraphics[width=1.0\linewidth]{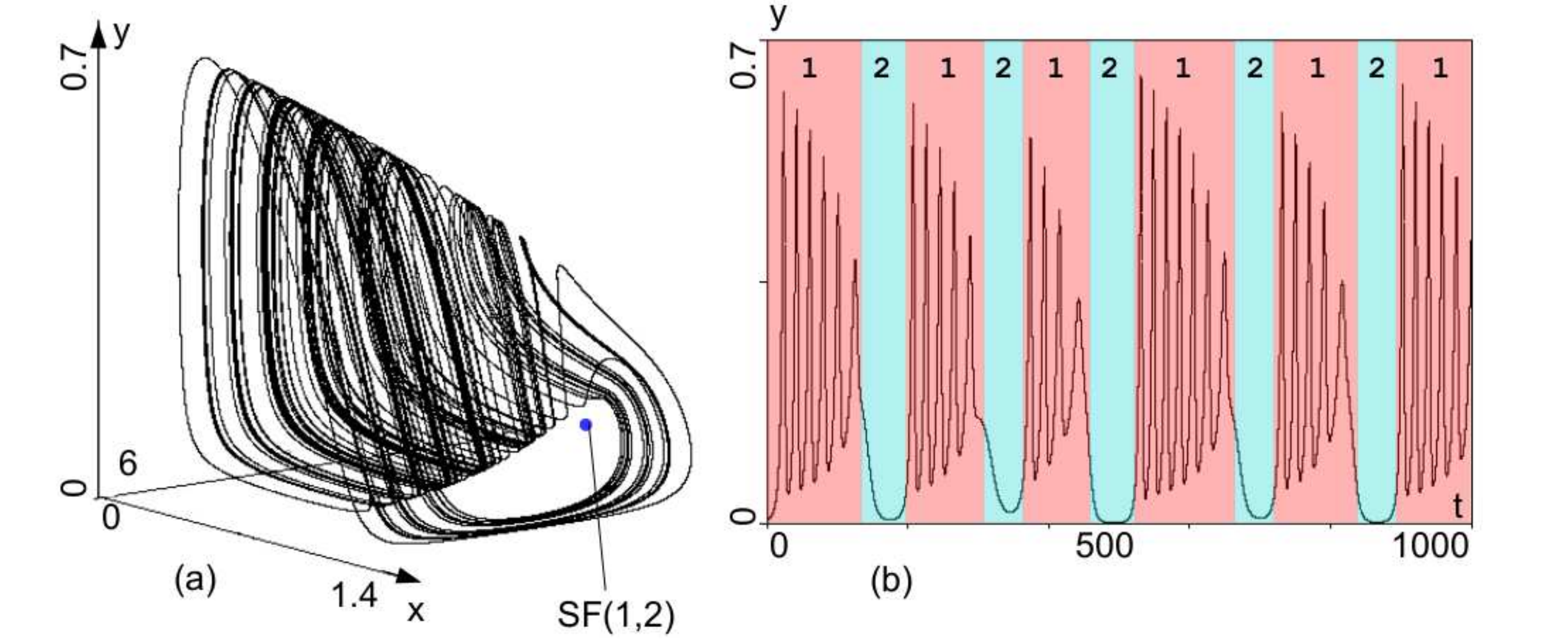} }
\caption{{\footnotesize Chaotic non-homoclinic bursting activity. (a) The phase portrait of the attractor and (b) time series $y(t)$ for ``teacup'' attractor at $r = 1.2, K = 1.063$ (the point $A$ in Fig. \ref{fig:LE_Diagram}).}}
\label{fig:chaotic_bursting}
\end{figure}

{\bf Type 1.}
{\it Chaotic non-homoclinic bursting activity.}
Such type of bursting for the system is generated by non-homoclinic multi-funnel (``teacup'') attractors, which appear in chaotic regions far from the homoclinic bifurcation curve, see e.g. the point $A$ in Fig. \ref{fig:LE_Diagram}. For this type of bursting random number of fast oscillations alternates with motions along the stable slow manifolds of the system, see Fig. \ref{fig:chaotic_bursting}b.

\begin{figure}[!ht]
\center{\includegraphics[width=1.0\linewidth]{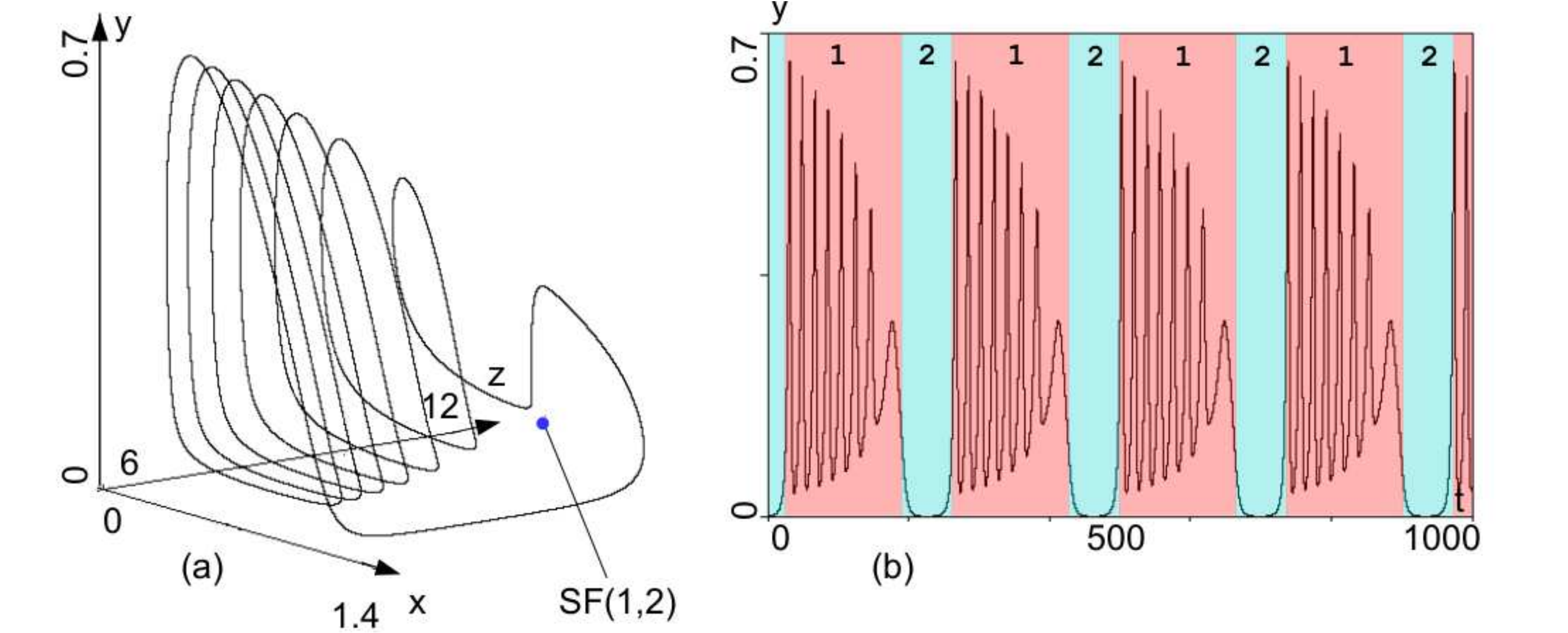} }
\caption{{\footnotesize Regular bursting activity. (a) The phase portrait of the attractor and (b) time series $y(t)$ for Regular bursting activity at $r = 1.2, K = 1.064$ (the point $B$ in Fig. \ref{fig:LE_Diagram}).}}
\label{fig:regular_bursting}
\end{figure}

{\bf Type 2.}
{\it Regular bursting activity.}
This type of bursting is generated by the stable limit cycles, which appear in the stability windows of non-homoclinic multi-funnel attractors e.g. in the point $B$, see Fig. \ref{fig:LE_Diagram}. Such limit cycles repeat orbits behavior of the corresponding attractors, and, as a rule, have the same number of turns as these attractors, see Fig. \ref{fig:regular_bursting}.

\begin{figure}[!ht]
\center{\includegraphics[width=1.0\linewidth]{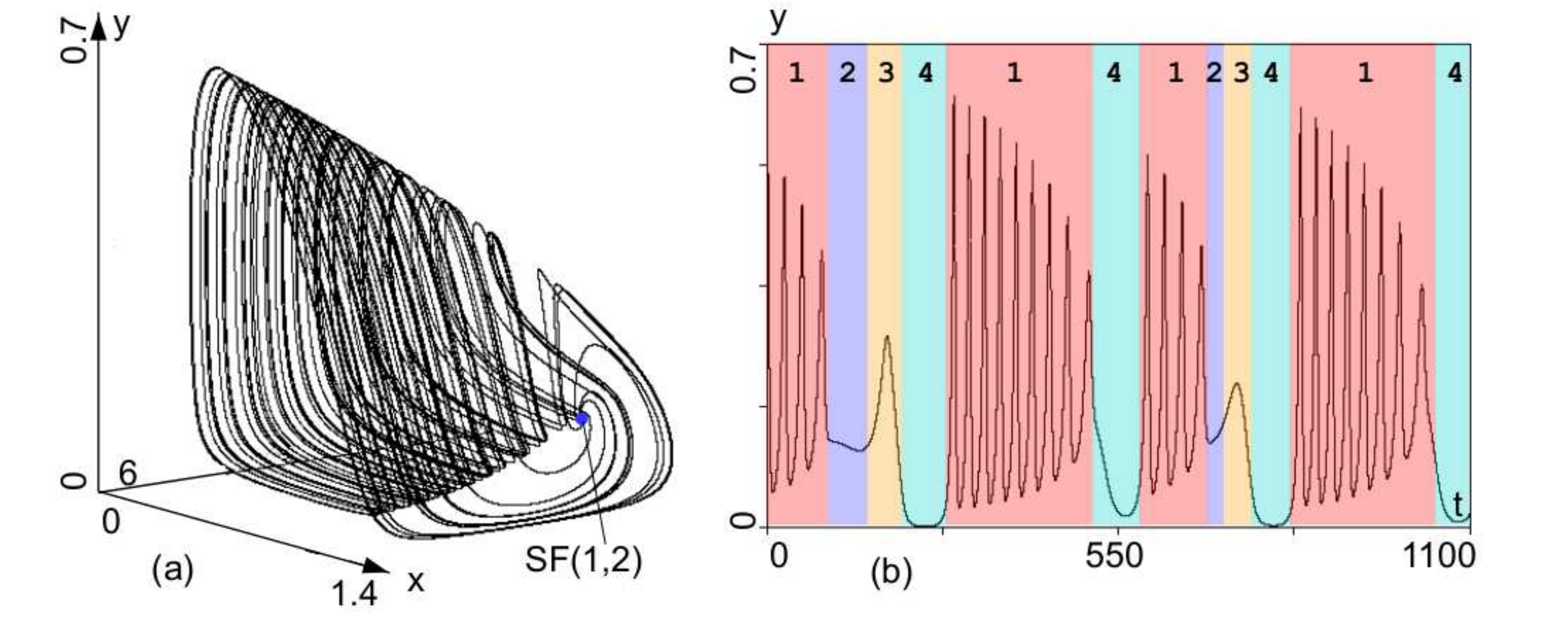} }
\caption{{\footnotesize Chaotic homoclinic bursting activity. (a) The phase portrait of the attractor and (b) time series $y(t)$ for ``teacup'' attractor at $r = 1.1, K = 1.09841$ (the point $C$ in Fig. \ref{fig:LE_Diagram}).}}
\label{fig:chaotic_bursting_3t}
\end{figure}

{\bf Type 3.}
{\it Chaotic homoclinic bursting activity.}
This type of bursting is associated with the co-existence of slow-fast behavior in the system with homoclinic multi-funnel attractors and, thus, appear in the chaotic regions along the homoclinic bifurcation curve $h$ e.g. in the point $C$, see Fig. \ref{fig:LE_Diagram}. The phase portrait of the attractor and time series in the point $C$ are presented in Fig. \ref{fig:chaotic_bursting_3t}.
As one can see, a random number of fast oscillations for this bursting (regions 1 in Fig. \ref{fig:chaotic_bursting_3t}b) alternates with two types of slow motions: small amplitude oscillations near a saddle-focus equilibrium (regions 2) and motions along the stable slow manifold of the fast subsystem (regions 4). It is worth noting that slow motions associated with orbits passing near the saddle-focus appear irregularly, namely when orbits, after fast oscillations, return to the vicinity of the saddle-focus. The closer an orbit passes near the saddle-focus, the longer time of the slow motions in region 2 is. Thus, in contrast to chaotic non-homoclinic bursting, time interval between bursts can be unbounded.
After the slow motions in regions 2 orbits oscillate with a small amplitude near the saddle-focus (regions 3) and then jump to the stable slow manifold.

\begin{figure}[!ht]
\center{\includegraphics[width=1.0\linewidth]{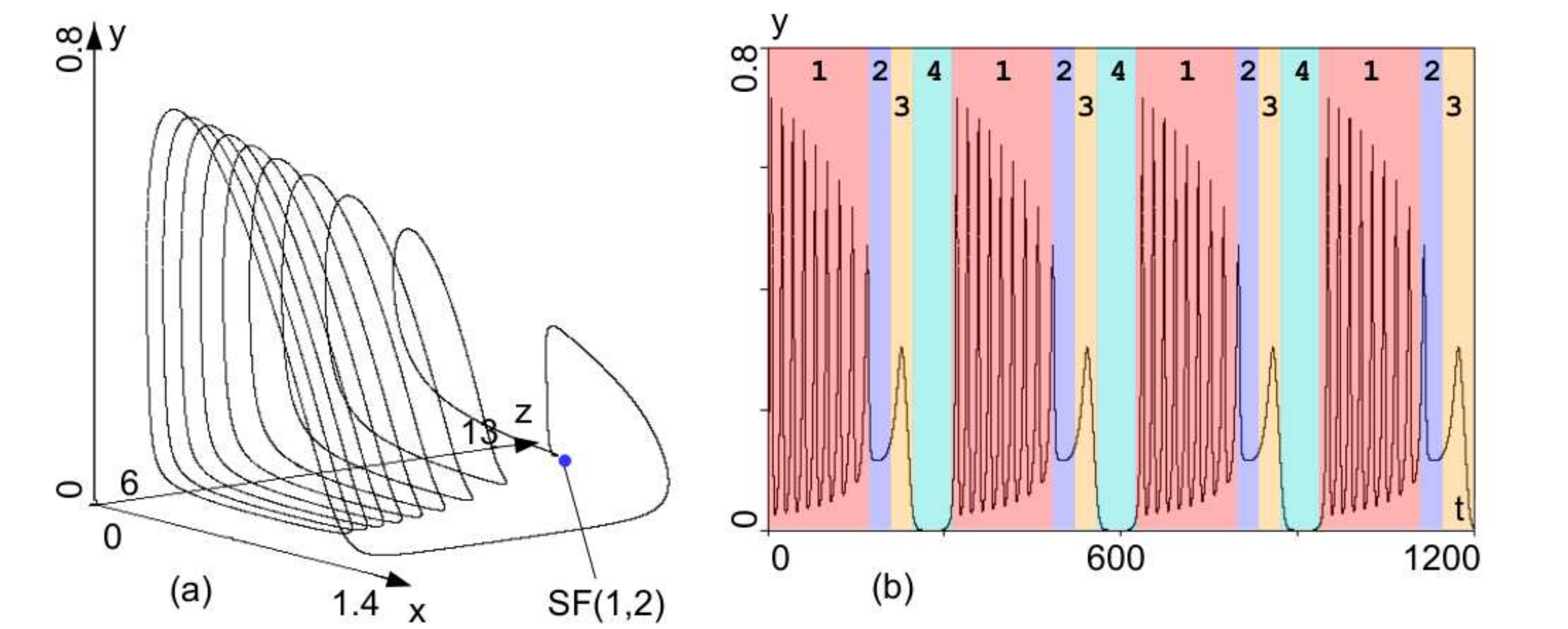} }
\caption{{\footnotesize Regular ``homoclinic'' bursting activity. (a) The phase portrait of the attractor and (b) time series $y(t)$ for ``teacup'' attractor at $r = 1.223, K = 1.1026$ (the point $D$ in Fig. \ref{fig:LE_Diagram}).}}
\label{fig:regular_bursting_3t}
\end{figure}

{\bf Type 4.}
{\it Regular ``homoclinic'' bursting activity.}
This type of bursting is generated on the multi-round limit cycles which come enough close to the saddle-focus equilibrium, e.g. in the point $D$, see Fig. \ref{fig:LE_Diagram}. It is well-known that spiral attractors belong to the class of quasiattractors, i.e. they either contain stable periodic orbits, or such orbits appear with an arbitrary small perturbation. As a rule, these periodic orbits (limit cycles) repeat the behavior of homoclinic loops in the attractors. In contrast to spiral attractors, limit cycles persist under small perturbations and exist in some open regions of the parameter space. Some limit cycles come close enough to the saddle focus equilibrium, see Fig. \ref{fig:regular_bursting_3t}a. The motions along the corresponding part of these cycles will be slow, see Fig. \ref{fig:regular_bursting_3t}b (region 2). Note, that for regular ``homoclinic'' bursting activity fast oscillations (with fixed number of spikes) alternate with regular slow motions near the saddle focus (regions 2) and with regular motions along the stable slow manifold of the fast subsystem (regions 4). The transition between regions 2 and 4 corresponds to the passing from vicinity of the saddle-focus equilibrium to the slow stable manifold (regions 3 in Fig. \ref{fig:regular_bursting_3t}b).

\section{Conclusion}

We demonstrated that the homoclinic spiral attractors in the Rosenzweig-MacArthur system are the root of two new types of bursting activity: chaotic homoclinic bursting activity corresponding to multi-funnel strange attractors, and regular ``homoclinic'' bursting activity corresponding to the stable periodic orbits occurring near the homoclinic attractors. For both these types of bursting fast oscillations alternate with slow motions of two kinds: near the saddle-focus and along the stable manifold of two dimension fast subsystem. In contrast to chaotic bursting, where fast oscillations have a different number of spikes and slow motions near the saddle-focus appear randomly, for regular bursting both fast oscillations and two types of slow motions are regular, i.e they have a constant time.

It is important to note, that the presented results can be transferred to such models as Hindmarsch-Rose system \cite{HindmarshRose1984}, chemical oscillator \cite{GaspardNicolis1983} and other systems which have a homoclinic orbit to the saddle equilibrium of (1,2) type and can be considered as a slow-fast systems with two-dimension fast subsystem and one-dimension slow subsystem.

{\bf Acknowledgement.}

The authors thank S.V. Gonchenko and A.L. Shilnikov for valuable advices and comments.
This paper was supported by RSF grant 17-11-01041 (Sections 2-4) and by RSF grant 14-12-00811 (Sections 5-6).
The authors also thanks RFBR grants 16-01-00364 and 16-32-00835 for the support of scientific research.

\end{document}